\documentclass[sigconf,nonacm]{acmart}
\setcopyright{none}   
\makeatletter
\renewcommand\@copyrightpermission{} 
\makeatother
\AtBeginDocument{%
  }

\copyrightyear{2025}
\acmYear{2025}
\setcopyright{rightsretained}
\acmConference[ASPDAC '25]{30th Asia and South Pacific Design Automation Conference}{January 20--23, 2025}{Tokyo, Japan}
\acmBooktitle{30th Asia and South Pacific Design Automation Conference (ASPDAC '25), January 20--23, 2025, Tokyo, Japan}\acmDOI{10.1145/3658617.3697777}
\acmISBN{979-8-4007-0635-6/25/01}

\usepackage{subcaption}
\usepackage{amsmath,amsfonts}
\usepackage{algorithmic}
\usepackage{algorithm}
\usepackage{comment}
\usepackage{xspace}
\usepackage{enumitem}

\begin{document}


\title{HDLFORGE: A Two-Stage Multi-Agent Framework for Efficient Verilog Code Generation with Adaptive Model Escalation}


\author{Armin Abdollahi}
\email{arminabd@usc.edu}
\orcid{0009-0007-1387-0995}
\affiliation{%
  \institution{University of Southern California}
  \city{Los Angeles}
  \state{CA}
  \country{USA}
}

\author{Saeid Shokoufa}
\email{shokoufa@usc.edu}
\affiliation{%
  \institution{University of Southern California}
  \city{Los Angeles}
  \state{CA}
  \country{USA}
}

\author{Negin Ashrafi}
\email{ashrafi@stanford.edu}
\orcid{0009-0003-8414-2996}
\affiliation{%
  \institution{Stanford University }
  \city{Stanford}
  \state{CA}
  \country{USA}
}

\author{Mehdi Kamal}
\email{mehdi.kamal@usc.edu}
\orcid{0000-0001-7098-6440}
\affiliation{%
  \institution{University of Southern California}
  \city{Los Angeles}
  \state{CA}
  \country{USA}
}

\author{Massoud Pedram}
\email{pedram@usc.edu}
\orcid{0000-0002-2677-7307}
\affiliation{%
  \institution{University of Southern California}
  \city{Los Angeles}
  \state{CA}
  \country{USA}
}

\renewcommand{\shortauthors}{Abdollahi et al.}

\begin{abstract}
\vspace{-2pt}
We present HDLFORGE, a two-stage multi-agent framework for automated Verilog generation that optimizes the trade-off between generation speed and accuracy. The system uses a compact coder with a medium-sized LLM by default (Stage A) and escalates to a stronger coder with an ultra-large LLM (Stage B) only when needed, guided by a calibrated score from inexpensive diagnostics including compilation, lint, and smoke tests. A key innovation is a counterexample-guided formal agent that converts bounded-model-checking traces into reusable micro-tests, significantly reducing bug detection time and repair iterations. The portable escalation controller can wrap existing Verilog LLM pipelines without modifying their internals. 
Evaluated on VerilogEval Human, VerilogEval V2, and RTLLM benchmarks, HDLFORGE demonstrates improved accuracy–latency trade-offs compared to single-stage systems through comprehensive analysis of wall-clock time distributions, escalation thresholds, and agent ablations.
On VerilogEval Human and VerilogEval V2, HDLFORGE-Qwen achieves 91.2\% and 91.8\% Pass@1 with roughly 50\% lower median latency, dramatically improving accuracy over other medium-sized models, and 97.2\% Pass@5 on RTLLM. 
\vspace{-10pt}
\end{abstract}




\maketitle

\section{Introduction}
The adoption of Large language models (LLMs) for code generation has grown substantially in HDL coding and hardware design, driven by their capability to automate tedious coding tasks, improve productivity, and enable designers to focus on high-level architectural decisions rather than low-level implementation details. However, while LLMs can generate Verilog RTL from natural language descriptions, they often produce code with syntax errors, functional bugs, and hallucinations, even for small designs \cite{zhang2025understanding}. 
Recent Verilog-LLM systems have adopted hardware-aware priors, tool support, and verification-driven repair approaches\cite{zhang2025understanding,zuo2025complexvcoder,wang2025insights,nadimi2024multi,yang2025haven}, leading to substantial improvements in generating correct HDL code as demonstrated on two prominent benchmarks: VerilogEval and RTLLM \cite{liu2023verilogeval,lu2024rtllm}. However, they typically fix the backbone model scale and treat wall-clock time and tool cost as secondary. Beyond direct RTL generation, recent work has also shown that LLMs can support other HDL-level hardware design tasks, such as code-level power, performance, and area estimation, further highlighting the broader value of language-model reasoning in hardware design workflows \cite{abdollahi2025rocketppa}. More broadly, AI-driven methods have also been explored across adjacent hardware-design settings, including accelerator design and optimization-oriented CAD workflows, underscoring that learning-based techniques are increasingly influencing the hardware stack beyond direct RTL generation \cite{abdollahi2024menage, abdollahi2025icd}. This mirrors trends in other domains where multi-stage and teacher-student architectures have improved prediction under resource constraints \cite{jin2025novel}, and where diagnostic signal fusion from heterogeneous data sources has driven accuracy gains \cite{abdollahi2025advanced} with similar data-driven frameworks scaling to large-scale network analysis \cite{sun2025optimizing}.

Multi-agent code generation frameworks demonstrate that decomposing complex coding tasks into specialized sub-tasks, where each agent assumes responsibility for one or more specific roles (e.g., coder and debugger), can significantly improve both robustness and sample efficiency. This collaborative approach enables agents to leverage their specialized capabilities, iterate on code through multiple refinement cycles, and catch errors that single-agent systems might overlook. By distributing responsibilities across multiple agents, these frameworks can handle larger and more complex codebases while maintaining higher code quality\cite{chatdev2023,islam2025codesim,islam2024mapcoder,khanzadeh2025agentmesh,zhao2025stackpilot,tao2024magis,ishibashi2024self,almorsi2024guided,sami2025nexus}. 

LLM-based Verilog generation has advanced through both multi-agent coordination and HDL-specific model tuning. On the systems side, MAGE, CoopetitiveV, and VerilogCoder use specialized agents or graph-based planning plus waveform tracing to iteratively generate, test, and debug RTL \cite{zhao2024mage,mi2024coopetitivev,ho2025verilogcoder}. On the model side, AutoVCoder, CodeV, and OriGen apply Verilog-focused pretraining, instruction tuning, and curated datasets to improve robustness under distribution shift \cite{gao2024autovcoder,zhao2025codev,cui2024origen}. These approaches substantially boost accuracy on VerilogEval and RTLLM but fix the backbone model scale and do not explicitly decide when to escalate to more capable (and more expensive) models. 

This paper introduces HDLFORGE, a two-stage multi-agent framework for Verilog generation that explicitly trades code generation latency against accuracy. Stage A uses a compact coder built on a medium-sized LLM and lightweight tool-driven feedback, while Stage B invokes a stronger coder backed by an ultra-large LLM only when needed. In contrast to prior LLM-for-Verilog systems, HDLFORGE integrates a counterexample-guided formal agent that turns bounded-model-checking traces into reusable micro-tests and couples it with a calibrated escalation score over inexpensive diagnostics (compile and lint results, smoke tests, trace stability, and remaining budget) to decide when to escalate from Stage A to Stage B; this specific combination of compact coder, large-model specialist, judge, and formal amplifier agents is, to our knowledge, new in the Verilog multi-agent literature. We evaluate HDLFORGE on VerilogEval Human, VerilogEval V2, and RTLLM \cite{liu2023verilogeval,lu2024rtllm}, and study wall-clock time distributions, escalation-threshold sweeps, and agent ablations to characterize accuracy–latency operating points. We make the following contributions:
\vspace{-0.3em}
\begin{itemize}[left=0pt]
    \item We introduce HDLFORGE, a two-stage multi-agent framework for Verilog generation that uses a small-model coder by default and escalates to a stronger coder only when a calibrated MG-Verilog score over inexpensive compile/lint/smoke diagnostics triggers it, improving the accuracy–latency trade-off over single-stage systems.
    \item We factor the method into a portable escalation controller that can wrap existing Verilog LLM pipelines such as AutoVCoder and VerilogCoder without modifying their internal prompts, retrieval, or tools, improving their speed–accuracy trade-off under a fixed backbone.
    \item We elevate a tiny formal amplifier agent into a CEGIS-style micro-test generator: counterexample traces are turned into micro-tests that substantially increase bug detection and reduce repair iterations and wall-clock time on a bug-injection benchmark.
    \item We structure HDLFORGE as a closed-loop multi-agent system in which a compact coder, a large specialist, a tool-driven judge, and a formal amplifier interact only through tool-level signals (scores, traces, and tests), yielding a design that, to our knowledge, has not been explored in prior LLM-for-Verilog frameworks.
\end{itemize}

\begin{figure*}[]
    \centering
    \includegraphics[width=0.9\linewidth]{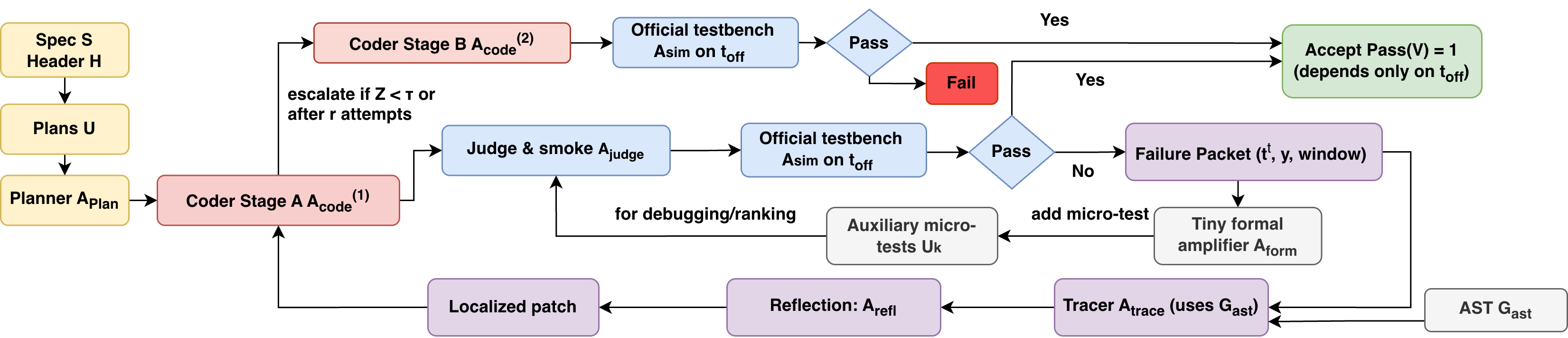}
    \vspace{-12pt}
    \caption{HDLFORGE two-stage cascade: Stage~A iterates generate--judge--repair with micro-tests, and escalates once to Stage~B when progress stalls; both stages are validated on the official testbench.}
    \label{fig:0}
\end{figure*}

\vspace{-1.0em}
\section{HDLFORGE Architecture}

In this section, we describe HDLFORGE, a two-stage cascade framework for efficient Verilog code generation that adaptively decides when to use expensive computational resources based on task difficulty. Its general archietcture is illustarted in Fig. \ref{fig:0}. 

\subsection{Problem Setting and Objective}

Each task consists of three inputs: a natural language specification $S$, a Verilog module header $H$ with port definitions, and an official testbench $t_{\mathrm{off}}$ that determines correctness. Our goal is to generate a Verilog implementation $V$ (based on the given tuple of ($S$,$H$)) that passes the official testbench while minimizing wall-clock time.

The key challenge is balancing the functionality correctness of the generated code against computational expense—using powerful models for every task is wasteful, while using only lightweight models sacrifices code correctness. HDLFORGE addresses this through an adaptive two-stage approach that escalates from cheap to expensive models only when necessary.

\subsection{Two-Stage Cascade Architecture}

HDLFORGE employs an agentic architecture comprising seven coordinated agents that work together to enable highly efficient HDL code generation. Some agents are LLM-based, while others are tool-based and exclusively execute predefined tools. The LLM-based agents, whose roles and interactions are described in the following subsection, can be mapped to LLMs of varying sizes. In this work, we propose a two-stage approach with distinct computational profiles for agent mapping. These stages are as follows:

\textbf{Stage A (Primary Solver):} Uses a mid-size model (e.g., Qwen-7B) to iteratively generate and repair code through tool-guided feedback loops. This stage attempts to solve tasks through multiple iterations. Note that a larger model can be used for this part to improve the efficiency of the model.

\textbf{Stage B (Final Attempt):} Deploys ultra-large cloud-based models (e.g., Claude sonnet 3.5) when Stage A shows insufficient progress, as determined by our escalation scoring system. This provides a final high-quality attempt with capabilities that differ from those of Stage A.

The system always starts with Stage A and escalates at most once to Stage B, ensuring bounded computational cost. 

\subsection{Stage A Core Workflow}

The HDLFORGE workflow proceeds through the following steps for each input task:

\subsubsection{Step 1: Multi-Plan Generation}
Given the specification $S$ and header $H$, an LLM-based Planner agent ($A_{\mathrm{plan}}$) generates $n$ alternative high-level implementation strategies:
\begin{equation}
\mathcal{P} = \{p_1, p_2, \dots, p_n\}.
\end{equation}
Each plan $p_j$ explains how to implement the module and summarizes the key invariants points by $S$. To obtain diverse plans from the same input $(S,H)$, we query the same planner multiple times using different decoding temperatures and sampling seeds, so that $\{p_j\}$ differ due to stochastic decoding rather than changes in the prompt or inputs.

\subsubsection{Step 2: Candidate Implementation}
For each plan $p_j$, the LLM-based Coder agent ($A_{\mathrm{code}}$) in Stage A generates multiple candidate implementations, creating a diverse set of initial solutions. This parallel exploration increases the likelihood of finding a correct implementation quickly.

\subsubsection{Step 3: Rapid Candidate Selection}
Given a batch of Verilog candidates that have been generated by the Coder agent ($A_{\mathrm{code}}$), the Judge and Smoke agent uses cheap “smoke” checks to keep only the most promising design. For each candidate $V$, the agent first runs compilation and linting with Verilator/Icarus and discards any design that fails to compile (e.g., due to a syntax error). For the remaining candidates, the agent runs
a short smoke simulation against the current micro-test set $U_k$, where $U_k$ denotes the auxiliary pool of tiny deterministic testbenches accumulated so far. Each $\mathrm{tb} \in U_k$ is simulated for at most $W_{\mathrm{smoke}}$ cycles, where $W_{\mathrm{smoke}}$ is the number of clock cycles for which each micro-testbench $\mathrm{tb} \in U_k$ is simulated during smoke testing, and the agent records the fraction of cycles where the candidate’s outputs match the expected values. The test set is initialized as $U_0$ with a tiny smoke checker derived automatically from $(S,H)$, e.g., a reset sequence followed by one or two representative input patterns. Note that the process of expanding $U_k$ beyond $U_0$ (via harness failures and formal counterexamples) is provided in Step~5, which is the task of another agent.

Candidates are ranked by a simple score (described in Section~\ref{adp}) that prefers successful compilation, higher micro-test pass fractions, and fewer lint warnings, and the top-ranked candidate is sent to Step~4 for full testing with the official testbench $t_{\mathrm{off}}$.

\subsubsection{Step 4: Full Testing and Failure Analysis}
The best candidate undergoes full testing with $t_{\mathrm{off}}$ in the simulation agent ($A_{sim}$). If it passes, we're done. Otherwise,  the failure should be analyzed, which is carried out by two agents:

\textbf{AST-Guided Backtracing:} The Tracer agent ($A_{\mathrm{trace}}$) constructs an Abstract Syntax Tree (AST) $G_{\mathrm{ast}}$ of the failing code. When a signal $y$ fails at time $t^\dagger$, the agent traverses the AST backwards from the failure point, collecting all signals and statements within depth $D_{\max}$ that could have contributed to the error. This creates a \emph{suspect cone} $S(u)$ pinpointing the likely bug location.

\textbf{Intelligent Repair Proposal:} A reflection agent ($A_{\mathrm{reft}}$) analyzes the failure context—including waveforms, the suspect cone, and plan invariants—to propose targeted fixes rather than regenerating entire modules.

\subsubsection{Step 5: Micro-Test Accumulation}

In parallel with repair, a tiny formal amplifier agent ($A_{\mathrm{form}}$) derives a lightweight set of safety properties from $(S,H)$. These properties encode generic reset, range, temporal, and state-encoding invariants that any acceptable implementation should satisfy, without committing to a particular realization strategy.

For each candidate $V$, $A_{\mathrm{form}}$ runs bounded model checking up to a time horizon of $d$ clock cycles. Within this $d$-cycle window, either all properties hold or a violation is exposed together with a counterexample trace $\tau_{\mathrm{ce}}$ that witnesses the failure. Each trace is then compiled into a deterministic micro-testbench $\mathrm{tb}(\tau_{\mathrm{ce}})$ that replays the finite input prefix in $\tau_{\mathrm{ce}}$ and checks the corresponding property outcome. The auxiliary test set is updated as
\begin{equation}
U_{k+1} \leftarrow U_k \cup \{\mathrm{tb}(\tau_{\mathrm{ce}})\}.
\end{equation}

These micro-tests act as a compact, executable summary of past formal failures: they are short (bounded by $d$), targeted to specific invariants, and inexpensive to run during smoke testing. Once a violation has been materialized as a micro-test, any future candidate that reintroduces the same flaw is rejected immediately by $U_{k+1}$, without requiring another formal query or a full pass of the official testbench.

\subsubsection{Step 6: Repair and Iteration}
Using the failure analysis, Stage A generates localized repairs focused on the suspect cone. The process returns to the coder agent, but now with enhanced micro-tests that provide rapid feedback on known issues.


\subsection{Adaptive Escalation Decision}
\label{adp}

After each Stage A attempt, HDLFORGE computes five diagnostic signals that indicate likelihood of success:
\vspace{-0.7 em}
\begin{equation}\label{eq:s_vector}
s = [s_{\mathrm{comp}}, s_{\mathrm{lint}}, s_{\mathrm{smoke}}, s_{\mathrm{trace}}, s_{\mathrm{budget}}] \in [0,1]^5
\end{equation}

The compile indicator is:
\begin{equation}\label{eq:s_comp}
s_{\mathrm{comp}} = \mathbf{1}\{\text{Verilator or Icarus build succeeds}\}
\end{equation}
Hence, $s_{\mathrm{comp}}=1$ when the design compiles successfully and $0$ otherwise.
The lint score measures code quality based on warning count, normalized using a maximum threshold $L_{\mathrm{lint}}$ (e.g., 30 warnings):
\begin{equation}\label{eq:s_lint}
s_{\mathrm{lint}} = 1 - \frac{\min(\ell, L_{\mathrm{lint}})}{L_{\mathrm{lint}}}
\end{equation}
where $\ell$ is the number of unique lint warnings. When $\ell=0$, $s_{\mathrm{lint}}=1$ (perfect quality); it decreases linearly as warnings increase, saturating at $0$ when $\ell \geq L_{\mathrm{lint}}$.
Smoke consistency averages the per-cycle match rate over a short smoke run of $W_{\mathrm{smoke}}$ cycles:
\begin{equation}\label{eq:s_smoke}
s_{\mathrm{smoke}} = \frac{1}{W_{\mathrm{smoke}}}\sum_{t=1}^{W_{\mathrm{smoke}}} \mathrm{match}(t)
\end{equation}
where $\mathrm{match}(t)=1$ if outputs agree at cycle $t$, and $0$ otherwise.

Trace stability measures the consistency of failure locations across attempts. Let $(y,t^\dagger)$ be the current failing signal and time, and $(y_{\mathrm{prev}}, t^\dagger_{\mathrm{prev}})$ be from the previous attempt. We save a waveform window of $\Delta t_{\mathrm{wave}}$ cycles (e.g., 64 cycles before and after the failure) for debugging. The stability score is:
\begin{align}
s_{\mathrm{trace}} = \tfrac{1}{2}\mathbf{1}\{y=y_{\mathrm{prev}}\} \nonumber
+ \tfrac{1}{2}\left(1-\frac{\min(|t^\dagger-t^\dagger_{\mathrm{prev}}|, \Delta t_{\mathrm{wave}})}{\Delta t_{\mathrm{wave}}}\right)
\label{eq:s_trace}
\end{align}
with $s_{\mathrm{trace}}=\tfrac{1}{2}$ if no prior failure exists. This rewards consistent failure patterns (same signal and nearby time), indicating Stage A is converging on the bug location.
Remaining Stage A budget tracks attempt usage:
\begin{equation}\label{eq:s_budget}
s_{\mathrm{budget}} = 1 - \frac{\text{attempts used}}{r}
\end{equation}
where $r$ is the maximum allowed Stage A attempts.
When a component is unavailable (e.g., smoke cannot run if compilation fails), its score defaults to zero, except $s_{\mathrm{trace}}=\tfrac{1}{2}$ initially.

The five signals combine into an escalation score:
\begin{equation}\label{eq:z_linear}
Z = w_0 + \sum_{i=1}^{5} w_i s_i
\end{equation}

The weights $\{w_i\}$ are fitted once on validation data using L2-regularized logistic regression to predict Stage A success probability, then calibrated via isotonic regression.
Given $Z$, the HDLFORGE flow controller escalates when:
\begin{equation}\label{eq:escalate_rule}
\text{Escalate to Stage B} \Longleftrightarrow (Z < \tau) \lor (\text{attempts} \geq r)
\end{equation}

The threshold $\tau$ controls the accuracy-efficiency tradeoff and is selected via validation sweep.

\subsection{Stage B: Informed Power Mode}

When escalation occurs, Stage~B is invoked with the original specification and header $(S,H)$, a summary of Stage~A failures and suspect cones, and the current micro-test set $U_k$. Using this context, it queries an ultra large cloud-based model (e.g., Claude~sonnet~3.5) for a single high-quality candidate, which its output is then evaluated by the $A_{sim}$ agent.

\subsection{Portable Controller Design}

Although HDLFORGE is implemented as a multi-agent cascade, the escalation logic can be factored into a portable decision layer that consumes only the diagnostic vector $s$ and outputs escalation decisions via $Z(s)$ and~\eqref{eq:escalate_rule}. Any external Verilog generator
\begin{equation}
\mathcal{G}_{\mathrm{ext}}(S,H)\mapsto V
\end{equation}
can be treated as a Stage~A coder agent black box: the generator remains responsible for proposing $V$, while the controller attaches compile, lint, and smoke checks to compute $s$, decides whether to request another Stage~A attempt, and escalates once to a Stage~B coder when $Z<\tau$ or the attempt cap $r$ is exhausted. In the Results section, we instantiate this interface around AutoVCoder and VerilogCoder and show that the same calibrated controller improves their speed–accuracy trade-off without re-tuning.

\section{Experimental Setup}

The coders ($A_{\mathrm{code}}$) follow a fixed design checklist. All registers use a synchronous active-high reset so that state is initialized to known values, sequential logic is written with non-blocking assignments, and combinational logic uses complete assignments with explicit defaults. Finite-state machines use explicit encodings with safe defaults. External inputs used for configuration are sampled only during their designated windows and are ignored otherwise, and outputs that are irrelevant outside certain phases are driven to valid constants rather than left unknown. The official benchmark masks unknown signals (X values) in the reference implementation but flags them as potential errors in the design under test when the corresponding reference bit is defined. Following this convention helps avoid false failures caused by X values and ensures the generated RTL aligns with the grading criteria.

We fix the hyperparameters to the values shown in Table~\ref{tab:hyperparams}. These values were tuned once on a validation split and then held constant across all reported experiments. They control the trade-off between Stage A's search breadth and depth versus how frequently tasks escalate to Stage B, and we analyze their impact through ablation studies in later sections. We report $\mathrm{pass@1}$ on VerilogEval-Human and VerilogEval-V2, and $\mathrm{pass@5}$ on RTLLM, using
\begin{equation}\label{eq:t}
\text{pass@}k \;=\; \mathbb{E}_{\text{problems}}\!\left[\,1 - \frac{\binom{f-c}{k}}{\binom{f}{k}}\,\right]
\end{equation}
as the pass@k estimator, where $f$ is the total number of trials per problem, $k$ is the number of trials chosen, and $c$ is the number of successful trials.

\begin{table}[htbp]
\centering
\renewcommand{\arraystretch}{0.8} 
\caption{Default hyperparameters of HDLFORGE framework.}
\vspace{-12pt}
\label{tab:hyperparams}
\footnotesize
\begin{tabular}{@{}l c p{0.58\columnwidth}@{}}
\hline
\textbf{Symbol} & \textbf{Default} & \textbf{Meaning} \\
\hline
$ n $ & $3$ & Plans per task \\
$ m $ & $4$ & Stage A candidates per plan \\
$ D_{max} $ & $5$ & Depth of AST graph \\
$ r $ & $5$ & Total Stage A attempts before escalation \\
$ d $ & $10$ & Depth for the bounded checker \\
$ W_{\mathrm{smoke}} $ & $100$ & Cycles in the smoke simulation \\
$ L_{\max} $ & $30$ & Lines in the suspect slice \\
$ \tau $ & $0.5$ & Escalation threshold on $Z$ \\
$ \Delta t_{\mathrm{wave}} $ & $64$ & Cycles around $ t^\dagger $ saved to the waveform \\
\hline
\end{tabular}
\end{table}

\begin{table}[htbp]
\renewcommand{\arraystretch}{0.8} 
\centering
\caption{Agents and their back-ends in the case of HDLFORGE-Qwen variant. LLMs are invoked only for Planner, Coder(s), and Reflexion; other agents are tool-based.}
\label{tab:agents_backends}
\small
\resizebox{\columnwidth}{!}{%
\begin{tabular}{@{}lll p{0.50\columnwidth}@{}}
\hline
\textbf{Agent} & \textbf{Role} & \textbf{Back-end type} & \textbf{Back-end instance} \\
\hline
$A_{\mathrm{plan}}$ & Planner & LLM & Qwen2.5-Coder-7B-Instruct\\
$A_{\mathrm{code}}^{(1)}$ & Coder Stage A & LLM & Qwen2.5-Coder-7B-Instruct \\
$A_{\mathrm{judge}}$ & Judge \& smoke & Tool & svlint, Verilator + iverilog\\
$A_{\mathrm{sim}}$ & Official testbench & Tool & Verilator C++ simulation of $t_{\mathrm{off}}$ \\
$A_{\mathrm{trace}}$ & Tracer & Tool & Pyverilog \\
$A_{\mathrm{refl}}$ & Reflexion & LLM & Qwen2.5-Coder-7B-Instruct\\
$A_{\mathrm{form}}$ & Tiny formal amplifier & Tool & SymbiYosys + Yosys-SMTBMC \\
$A_{\mathrm{code}}^{(2)}$ & Coder Stage B & LLM & Claude 3.5 \\
\hline
\end{tabular}}
\end{table}

We evaluate two cascade variants that differ only in the choice of language model for the Stage-A coder agent, while keeping all other agents and hyperparameters fixed. The first variant uses a medium-sized model in Stage A (Qwen2.5-Coder-7B-Instruct) and an ultra-large model in Stage B (Claude~3.5); we refer to this configuration as HDLFORGE-Qwen. To test the portability of our controller and to assess whether escalation remains beneficial when Stage A is already strong, the second variant replaces the Stage-A model with a larger backbone (GPT-4o) while retaining Claude~3.5 in Stage B; we denote this configuration as HDLFORGE-GPT4o. The details of HDLFORGE-Qwen are provided in Table~\ref{tab:agents_backends}.

For wall-clock time, we evaluated HDLFORGE, MAGE, and VerilogCoder on the same workstation, an NVIDIA RTX A6000 (48\,GB), a 16-core CPU at \(\sim\)3.5\,GHz, 128\,GB RAM, and NVMe SSD storage. Because MAGE and VerilogCoder do not report timing in their papers, we executed their open-source implementations from GitHub under this setup. CoopetitiveV has neither released code nor reported timing, so it is omitted from wall-time comparisons.


We tune all hyperparameters and the escalation calibrator on 200 uniformly sampled MG-Verilog~\cite{zhang2024mg} training instances using their natural language descriptions and reference RTL, then freeze all controller settings. To prevent leakage into VerilogEval Human, VerilogEval V2, and RTLLM, we enforce strict disjointness by canonicalizing and hashing candidates, screening near duplicates, and dropping any item whose module interface matches a test task.

Among the controller hyperparameters, we expose the escalation threshold $\tau$ as the primary user-facing knob because it directly trades accuracy against latency. Other parameters such as the attempt cap $r$, the smoke-testing budget $W_{\mathrm{smoke}}$, the batch size $B$, and the BMC depth $d$ act mainly as resource limits rather than shaping the accuracy–latency frontier.

Table~\ref{tab:tau_sweep_datasets} shows the accuracy–time tradeoff governed by $\tau$. At $\tau=0.70$ the system is quickest with 55.2 seconds mean time but yields the lowest accuracy at 98.6\% MG-Verilog. Raising escalation at $\tau=0.50$ lifts accuracy to 99.8\%  while mean time increases to 74.0 seconds. Lowering to $\tau=0.30$ brings only marginal accuracy gains and a large time cost of 94.6 seconds. Hence $\tau=0.50$ is the most balanced setting in our validation.

\subsection{External Pipelines: AutoVCoder and VerilogCoder}
\label{sec:external_pipelines}

To evaluate HDLFORGE as a portable controller, we wrap two existing Verilog generation systems, AutoVCoder and VerilogCoder, as external Stage~A generators. In both cases we compare a \emph{baseline} setting, where the original system is run as released, with a \emph{controller-wrapped} setting, where the same system is treated as a black-box Stage~A under our escalation policy while keeping its backbone model, prompts, retrieval components, and training data unchanged.

\begin{table}[htbp]
\renewcommand{\arraystretch}{0.5} 
\centering
\caption{Escalation threshold sweep on MG-Verilog accuracy and mean time}
\vspace{-10pt}
\label{tab:tau_sweep_datasets}
\small
\setlength{\tabcolsep}{5pt}
\renewcommand{\arraystretch}{1.2}
\begin{tabular}{lccc}
\hline
& \multicolumn{3}{c}{$\tau$} \\
\cline{2-4}
Metric & 0.70 & 0.50 & 0.30 \\
\hline
MG-Verilog Pass@1 (\%)           & 98.6 & \textbf{99.8} & 99.8 \\
Mean TTP on MG-Verilog (s)    & 55.2 & \textbf{74.0} & 94.6 \\
\hline
\end{tabular}
\end{table}

In the baseline setting, the AutoVCoder is used in its released AutoVCoder--CodeQwen configuration (CodeQwen1.5-7B backbone) and evaluated on the VerilogEval-EvalMachine, VerilogEval-Human, and RTLLM, where we record Pass@1, Pass@5, and mean time-to-pass. VerilogCoder is used in its VerilogCoder--Llama3 configuration on VerilogEval-Human v2 with its original agent configuration, and we record Pass@1 and time-to-pass.

In the controller-wrapped setting, both systems are exposed to HDLFORGE through the same interface where given a task $(S,H)$, the external generator plays the role of Stage~A and proposes one or more candidate implementations, while the controller attaches compile, lint, smoke, and budget checks to produce the diagnostic vector $s$, computes $Z(s)$ via~\eqref{eq:z_linear}, and either accepts a Stage~A candidate, requests another Stage~A attempt (if available), or escalates once to Stage~B according to~\eqref{eq:escalate_rule}. For AutoVCoder, Stage~A may sample up to $r$ candidates per task under its native decoding and RAG pipeline. For VerilogCoder, a single VerilogCoder run yields one candidate and its pass/fail status on the official testbench, from which we derive $s$ by combining compile and lint outcomes with simulator mismatch statistics; we set $s_{\mathrm{trace}}$ and $s_{\mathrm{budget}}$ based on whether tracing fired and whether VerilogCoder exhausted its internal budget. If the resulting candidate passes and $Z \ge \tau$, the controller accepts it; otherwise it escalates once to Stage~B (Claude~3.5), which synthesizes a new candidate conditioned on the original specification and summarized failure information. This construction preserves the internal logic of both AutoVCoder and VerilogCoder while exposing them through a single, reusable controller layer.

\subsection{Bug-Injection Benchmark for Micro-Test Amplifier}
\label{sec:bug_injection_setup}

To isolate the effect of the micro-test amplifier, we construct a controlled bug-injection benchmark starting from known-correct Verilog modules. We select a subset of VerilogEval and RTLLM tasks whose reference implementations pass the official testbench and satisfy basic structural checks (no inferred latches, explicit reset, complete combinational assignments). For each reference design, we generate multiple buggy variants by injecting exactly one synthetic fault drawn from four classes: (i) off-by-one and boundary errors in counters and index bounds, (ii) reset bugs such as missing or incorrect initialization of state and output registers, (iii) finite-state-machine (FSM) bugs such as deadlocks, missing transitions, or unreachable states, and (iv) temporal and race bugs such as incorrect ordering of non-blocking assignments or missing enable conditions.

Each buggy instance then serves as input to three configurations under a shared maximum number of repair iterations and an identical overall tool-call budget. In the HDLFORGE + micro-tests configuration, the full system runs with the tiny formal amplifier $A_{\mathrm{form}}$ enabled, where in each repair iteration we mine properties and run bounded model checking up to depth $d$, and whenever a counterexample $\tau_{\mathrm{ce}}$ is found we synthesize a deterministic micro-testbench $\mathrm{tb}(\tau_{\mathrm{ce}})$ and append it to the auxiliary set $U_k$. In the HDLFORGE without micro-tests configuration, the controller, agents, and toolchain are identical, but $A_{\mathrm{form}}$ is disabled and $U_k = \varnothing$, so only the official testbench $t_{\mathrm{off}}$ is used for smoke testing and acceptance. In the \emph{external baseline} configuration, we run a representative existing Verilog LLM system (AutoVCoder--CodeQwen on VerilogEval and VerilogCoder--Llama3 on VerilogEval-Human v2) with its native verification mechanisms and no synthesized micro-tests.

For each buggy instance we run all configurations under the same fixed limits on the number of synthesis–feedback–repair iterations and the total number of tool calls (compilation, simulation, and formal checks). We then record (i) the bug detection rate, defined as whether at least one failing test (either the official testbench or a micro-test) is observed before these limits are reached, (ii) the number of iterations required to recover a correct implementation, and (iii) the wall-clock time. In the Results section we aggregate these metrics across bug types and configurations.

\section{Results}

\subsection{Accuracy}
Table~\ref{tab:ve_human_v2_with_models} reports Pass@1 on VerilogEval Human and VerilogEval V2, and Pass@5 on RTLLM, together with the underlying base LLMs for each system. Our primary configuration, HDLFORGE-Qwen, uses Qwen2.5-Coder-7B as the Stage-A coder and invokes Claude~3.5 only once per episode as a Stage-B specialist. Despite relying on a 7B base model, HDLFORGE-Qwen attains 91.2/91.8 Pass@1 on VerilogEval Human/V2 and 97.2 Pass@5 on RTLLM, dramatically improving over other 7B-based systems such as AutoVCoder (48.5 on VE-Human, 51.7 on RTLLM@5), CodeV (53.2/62.1), and OriGen (54.4/65.5). These gains indicate that the multi-agent design and micro-test/CEGIS feedback contribute substantially beyond simply choosing a stronger 7B backbone.

We also evaluate a higher-cost variant, HDLFORGE-GPT4o, which upgrades the Stage-A coder to GPT-4o while still using Claude~3.5 once as Stage B. HDLFORGE-GPT4o achieves the highest absolute scores in Table~\ref{tab:ve_human_v2_with_models}—95.5/96.8 Pass@1 on the VerilogEval Human/V2 and 99.8 Pass@5 on RTLLM— outperforming CoopetitiveV (94.9/96.0/99.8) and MAGE (94.8/95.7) that rely on repeated calls to a large Claude Sonnet backbone. The consistent margins across both VerilogEval splits suggest that HDLFORGE’s escalation controller and agent configuration yield more reliable generalization than prior multi-agent Verilog systems, even when all methods are built on similar large LLMs.

\vspace{-1 em}
\begin{table}[htbp]
\centering
\renewcommand{\arraystretch}{0.7} 
\caption{Pass@1 (\%) on VerilogEval Human and V2, and Pass@5 (\%) on RTLLM, with the underlying base LLM(s).}
\vspace{-12pt}
\label{tab:ve_human_v2_with_models}
\resizebox{\columnwidth}{!}{%
\begin{tabular}{lcccc}
\hline
\textbf{System} & \textbf{Base LLM} & \textbf{VE-Human} & \textbf{VE-V2} & \textbf{RTLLM@5} \\
\hline
HDLFORGE-GPT4o & GPT\;4o,\;Claude\;3.5 & 95.5 & 96.8 & 99.8 \\
HDLFORGE-Qwen  & Qwen2.5-Coder-7B,\;Claude\;3.5 & 91.2 & 91.8 & 97.2 \\
CoopetitiveV   & Claude\;3.5\;Sonnet & 94.9 & 96.0 & 99.8 \\
MAGE           & Claude\;3.5\;Sonnet & 94.8 & 95.7 & N/A \\
VerilogCoder   & GPT\;4\;Turbo       & N/A  & 94.2 & N/A \\
AutoVCoder     & CodeQwen1.5\;7B     & 48.5 & N/A  & 51.7 \\
CodeV          & Qwen2.5\!-\!Coder\;7B & 53.2 & N/A & 62.1 \\
OriGen         & DeepSeek\!-\!Coder\;7B & 54.4 & N/A & 65.5 \\
\hline
\end{tabular}}
\vspace{-1 em}
\end{table}

\subsection{Controller Portability on External Pipelines}
\label{sec:controller_portability_results}

Table~\ref{tab:controller_portability} evaluates HDLFORGE as a portable controller around AutoVCoder and VerilogCoder. For each external pipeline and dataset we compare the baseline system, run as originally released, with the controller-wrapped variant that treats the pipeline as Stage~A under our calibrated escalation policy. Backbone LLMs are kept identical within each pair (CodeQwen1.5-7B for AutoVCoder; Llama3-70B for VerilogCoder), and Stage~B uses Claude~3.5 only when the controller elects to escalate.

Across the VerilogEval-EvalMachine, VerilogEval-Human, and RTLLM, attaching HDLFORGE’s controller consistently improves Pass@1 by 3--5\,pp over AutoVCoder while increasing mean time-to-pass by less than 10\,\%. On VerilogCoder, the controller raises Pass@1 on VerilogEval-Human v2 by roughly 4.5\,pp with a modest wall-clock time overhead. Escalations remain relatively rare ($\leq 15\%$ of tasks), indicating that most instances are still handled by the original pipeline, with Stage~B reserved for genuinely hard problems. These results support the view of HDLFORGE as a reusable decision layer that can sit on top of existing Verilog LLM systems and improve their speed–accuracy trade-off without per-dataset re-tuning.

\vspace{-5pt}
\begin{table}[htbp]
\renewcommand{\arraystretch}{0.8} 
\centering
\caption{Controller portability: HDLFORGE wrapped around AutoVCoder and VerilogCoder. Time is median per-attempt latency, not time-to-first-pass.}
\vspace{-12pt}
\label{tab:controller_portability}
\scriptsize
\setlength{\tabcolsep}{3pt}
\resizebox{\columnwidth}{!}{%
\begin{tabular}{l l l c c c c}
\hline
Pipeline & Dataset & Ctrl. & Pass@1 & Pass@5 & Time [s] & Escal. \% \\
\hline
AutoVCoder--CodeQwen & EvalMachine & base & 69.0 & 79.3 & 18.2 & --   \\
                      &             & +HDF & 72.4 & 81.0 & 19.5 & 11.3 \\
\cline{2-7}
                      & EvalHuman   & base & 46.9 & 53.7 & 21.0 & --   \\
                      &             & +HDF & 51.5 & 58.2 & 22.7 & 14.8 \\
\cline{2-7}
                      & RTLLM       & base & --   & 51.7 & 24.6 & --   \\
                      &             & +HDF & --   & 56.9 & 25.3 & 9.7  \\
\hline
VerilogCoder--Llama3  & VE-Human v2 & base & 67.3 & --   & 46.8 & --   \\
                      &             & +HDF & 71.9 & --   & 49.1 & 12.5 \\
\hline
\end{tabular}%
}
\end{table}

\vspace{-12pt}
\subsection{Wall-clock time}

\begin{figure}[h]
    \centering
    \includegraphics[width=0.7\linewidth]{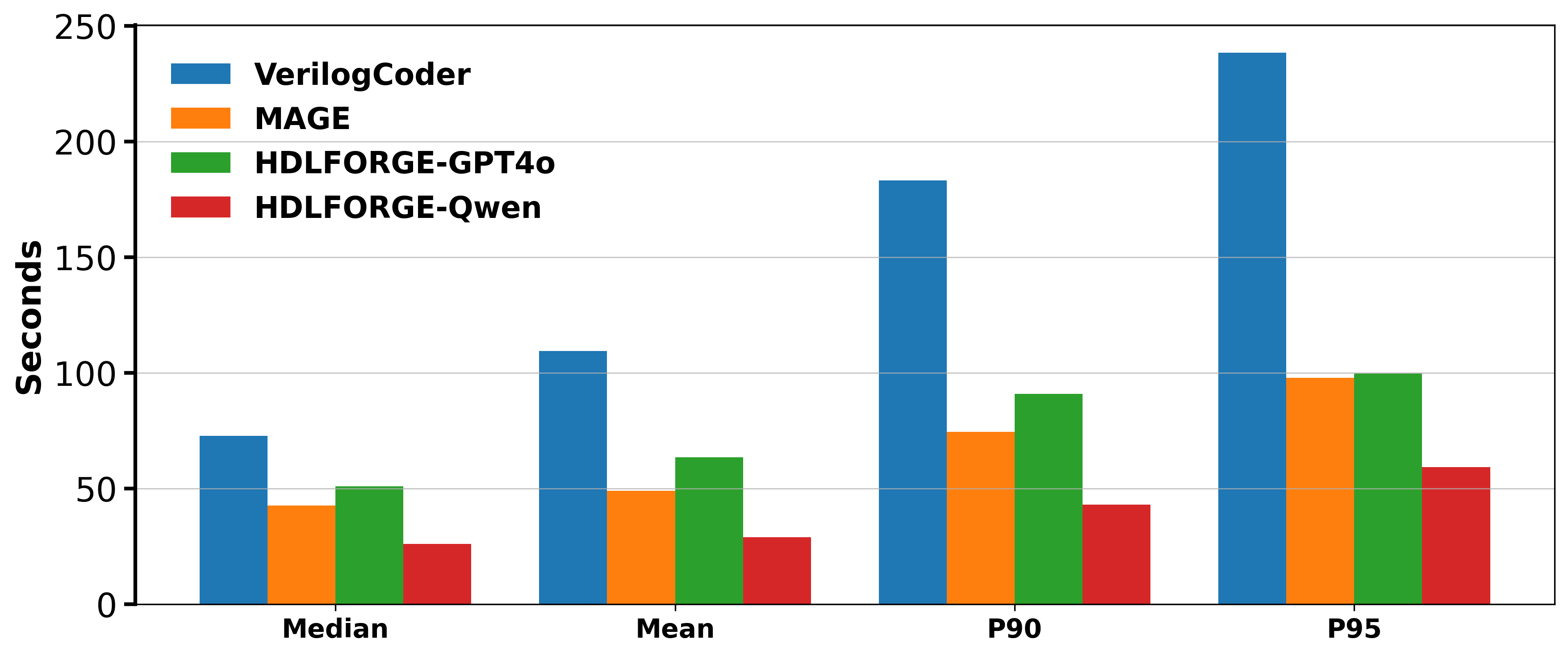}
    \vspace{-12pt}
    \caption{Time to pass summary for Stage A Stage B and total. Bars show median mean p90 and p95 on VerilogEval}
    \label{fig:2}
    \vspace{-2 em}
\end{figure}

Figure~\ref{fig:2} shows that HDLFORGE-Qwen is the fastest across all statistics with the lowest median and the smallest p90 and p95 tails, making it the preferred choice when rapid turnaround is the primary goal and with only a small accuracy drop relative to HDLFORGE-GPT4o. HDLFORGE-GPT4o targets higher accuracy and still maintains competitive latency with a much shorter tail than VerilogCoder and with medians that are close to MAGE. Thus, the speed-oriented HDLFORGE-Qwen is truly fast, and the accuracy-oriented HDLFORGE-GPT4o is strong with acceptable speed because it delivers high accuracy without incurring a large time penalty.

\subsection{Ablation Study}

Table~\ref{tab:agent_all_ablate} evaluates ablations on VerilogEval Human with the HDLFORGE GPT4o configuration at each variant’s best attainable operating point. Although removing an agent might seem to simplify the pipeline and reduce latency, every removal lowers Pass@1 by about four to five percentage points while also slowing convergence. Removing the judge and smoke stage hurts selection quality and lifts median time to 79.1 seconds. Removing the tracer weakens localization and yields 76.9 seconds. Withholding the reflexion note produces the largest accuracy drop and a median of 82.8 seconds. Disabling micro tests gives the slowest median at 84.1 seconds and still falls short in accuracy. The full system remains both more accurate at 95.5 percent Pass@1 and more time efficient at 64.0 seconds which shows that each agent materially improves effectiveness and efficiency rather than adding redundant overhead. We see the same trend on HDLFORGE-Qwen model.

\begin{table}[htbp]
\centering
\renewcommand{\arraystretch}{0.8} 
\caption{Agent-wise ablation on VerilogEval Human with HDLFORGE-GPT4o.}
\label{tab:agent_all_ablate}
\small
\resizebox{\columnwidth}{!}{%
\begin{tabular}{@{}l c c c c@{}}
\hline
\textbf{Configuration} & \textbf{Pass@1 (\%)} & $\boldsymbol{\Delta}$\textbf{pp} & \textbf{Median TTP (s)} & $\boldsymbol{\Delta}$\textbf{s} \\
\hline
Full system                      & 95.5 & ---    & 64.0 & ---   \\
w/o Judge \& smoke               & 90.8 & $-4.7$ & 79.1  & $+15.1$ \\
w/o Tracer ($G_{\mathrm{ast}}$)  & 91.8 & $-3.7$ & 76.9 & $+12.9$ \\
w/o Reflexion note               & 90.5 & $-5.0$ & 82.8 & $+18.8$ \\
w/o Micro-tests $U_k$            & 90.9 & $-4.6$ & 84.1 & $+20.1$ \\
\hline
\end{tabular}}
\end{table}

\subsection{Bug-Injection Study: CEGIS-Style Micro-Test Amplifier}
\label{sec:bug_injection_results}

We next quantify the impact of the micro-test amplifier in the controlled bug-injection benchmark described earlier. Table~\ref{tab:bug_injection_overall} reports overall detection and repair statistics aggregated over 200 buggy instances (50 per bug type). Both HDLFORGE variants substantially improve bug detection compared with running AutoVCoder or VerilogCoder in their native configurations, and the full HDLFORGE + micro-tests system further reduces the median number of repair iterations and wall-clock time despite the additional bounded-checking overhead.

\begin{table}[htbp]
\renewcommand{\arraystretch}{0.7} 
\centering
\caption{Bug-injection benchmark: overall detection and repair statistics.}
\vspace{-12pt}
\label{tab:bug_injection_overall}
\footnotesize
\setlength{\tabcolsep}{3pt}
\resizebox{\columnwidth}{!}{%
\begin{tabular}{l c c c}
\hline
Method                      & Detect. [\%] & Med. iters & Time [s] \\
\hline
AutoVCoder baseline         & 64.0         & 7.0        & 39.8     \\
VerilogCoder baseline       & 72.0         & 7.0        & 42.6     \\
HDLFORGE -- micro-tests     & 82.5         & 5.0        & 36.8     \\
HDLFORGE + micro-tests      & 95.0         & 3.0        & 33.1     \\
\hline
\end{tabular}%
}
\end{table}
\vspace{-10pt}

Table~\ref{tab:bug_injection_by_type} breaks down detection rates by bug type. The largest gains from micro-tests occur for reset and FSM bugs, where violations naturally manifest as short counterexample traces that can be encoded into deterministic micro-testbenches. Without micro-tests, both AutoVCoder and VerilogCoder frequently converge to implementations that pass the official testbench but still violate basic safety properties; once a counterexample is turned into a micro-test, HDLFORGE is forced to repair the underlying bug rather than overfitting to the limited official stimulus.

\vspace{-1 em}
\begin{table}[htbp]
\centering
\renewcommand{\arraystretch}{0.8} 
\caption{Bug-injection benchmark: detection rate by bug type.}
\vspace{-12pt}
\label{tab:bug_injection_by_type}
\footnotesize
\setlength{\tabcolsep}{3pt}
\resizebox{\columnwidth}{!}{%
\begin{tabular}{l c c c c}
\hline
Bug type         & AutoVCoder [\%] & VerilogCoder [\%] & HDF -- micro [\%] & HDF + micro [\%] \\
\hline
Off-by-one       & 78.0            & 86.0              & 90.0              & 96.0             \\
Reset bugs       & 50.0            & 58.0              & 76.0              & 96.0             \\
FSM bugs         & 56.0            & 64.0              & 80.0              & 94.0             \\
Temporal / race  & 72.0            & 80.0              & 84.0              & 94.0             \\
\hline
\end{tabular}%
}
\vspace{-2 em}
\end{table}

\section{Conclusion}

We presented HDLFORGE, a two-stage multi-agent framework for Verilog generation that explicitly optimizes accuracy–latency trade-offs through calibrated escalation between a compact coder and a large-model specialist. By integrating a counterexample-guided formal agent that transforms BMC traces into reusable micro-tests and coupling it with inexpensive diagnostic signals, HDLFORGE achieves improved operating points over single-stage baselines. The portable escalation controller wraps existing pipelines without internal modifications, making the approach broadly applicable.

\bibliographystyle{unsrt}
\bibliography{myref}

\end{document}